\pdfoutput=1 
\documentclass[a4paper, 11pt]{article}
\usepackage[dvipsnames]{xcolor}

\usepackage{jheppub}

\makeatletter
\gdef\@fpheader{}
\makeatother

\usepackage[T1]{fontenc}
\usepackage{amsmath,amssymb,amsthm,amsfonts,graphicx,hyperref}
\usepackage{stmaryrd}
\usepackage[utf8]{inputenc}
\usepackage{bm}

\hypersetup{ linktoc=all,
    colorlinks, linkcolor=blue,
    citecolor=magenta, urlcolor=red
}

\newcommand{\be}{\begin{equation}}
\newcommand{\ee}{\end{equation}}
\newcommand{\bea}{\begin{eqnarray}}
\newcommand{\eea}{\end{eqnarray}}

\makeatletter \@addtoreset{equation}{section}

\usepackage[most]{tcolorbox}
\tcbset{highlight math style={left=02mm,right=02mm,top=02mm,bottom=02mm}}
\usepackage{empheq}

\begin{document}

\title{\LARGE{\centerline{Null Strings Gauged and Reloaded, I:}} 
\centerline{Null Strings Have Carroll-Weyl Gauge Symmetry}}

\author[a]{M.M. Sheikh-Jabbari}
\author[b]{, H. Yavartanoo}
\affiliation{$^a$ School of Physics, Institute for Research in Fundamental
Sciences (IPM), P.O.Box 19395-5531, Tehran, Iran}
\affiliation{$^b$ Beijing Institute of Mathematical Sciences and Applications (BIMSA), Huairou District, Beijing 101408, P. R. China}
\emailAdd{  
jabbari@theory.ipm.ac.ir, yavar@bimsa.cn}

\abstract{Null strings, strings with  Carrollian worldsheets, are traditionally described by the Isberg-Lindstr\"om-Sundborg-Theodoridis (ILST) action, which is obtained via a tensionless limit  of standard tensile strings. In a recent work \cite{Sheikh-Jabbari:2026cnj}, we observed that the ILST action enjoys an overlooked partial-gauge symmetry whose existence calls into question the consistency of {standard} null-string analyses found in the literature.  In this paper, we show that the Carrollian geometry provides us with two Weyl scaling options, in contrast to a single Weyl scaling available for the ordinary tensile string worldsheet. 
Defining the null string theory by the action that realizes the two Carroll-Weyl scalings as well as the 2D diffeomorphisms as local (gauge) symmetries, we construct the new null string action. We show that the ILST action is obtained after fixing one of the two Carroll-Weyl scalings of the action that we construct, and that the residual part of this symmetry is precisely the overlooked partial-gauge symmetry discussed in \cite{Sheikh-Jabbari:2026cnj}. We hence clarify the Carroll-geometric origin of the overlooked symmetry and pave the way for a consistent quantization of null strings.}  
\maketitle
\flushbottom
\section{Introduction}
\label{sec:motivation_generalized_action}

Physical theories generically come with some parameters or constants of nature and a standard sanity check is whether in the special places/regions of their parameter space they recover or go over to the expected results. A prime example is the Galilean $c\to\infty$ limit ($c$ being the speed of light) of Lorentz invariant theories. Recently, physicists have shown interest in the opposite $c\to 0$ {limit, i.e the} ``Carrollian'' limit of relativistic theories, see e.g. \cite{Duval:2014uva, Henneaux:2021yzg, deBoer:2023fnj}. Various aspects of Carrollian (field) theories have been discussed and many others are still underway, see \cite{Bagchi:2025vri} and references therein. An important feature one should keep in mind in defining theories through parameter space limit of known theories, including the Carrollian theories, is that even when the limit is well-defined, the theory obtained in the limit may exhibit new features, e.g. enhanced (gauge) symmetries, that have no counterpart in the original theory.  In such cases, while one may obtain the action of the theory as a limit, the same limit procedure applied to analysis and results may yield incorrect/inconsistent results for the theory obtained through the limit. In this paper we discuss a particular instance of such symmetry enhancement, for which the limit {directly on the results of original theory is not valid.} 

String theory is defined through the Polyakov action as a 2D worldsheet theory with 2D diffeomorphism $+$ Weyl as local gauge symmetry \cite{Green:1987sp, Polchinski:1998rq}. The worldsheet of the ``tensile string'' is a 2D Lorentzian manifold and has a fundamental scale, string tension $T$, and a parameter, the string coupling $g_s$. The $T\to \infty$ limit (low energy limit) of string theory is a very much discussed limit where string theory is supposed to recover the standard physics described by quantum field theories (QFTs) \cite{Polchinski:1998rq, Polchinski:1998rr}. The opposite limit, the tensionless $T\to 0$ limit, is less discussed in the literature. It is a limit where stringy features are very pronounced and we are far from local QFT validity regime. The tensionless strings have arisen in two different contexts:\footnote{We thank Bo Sundborg and Giulio Bonelli for discussion on this point.} (1) When the background comes with a length scale, like AdS backgrounds and the tensionless limit corresponds to cases were curvature of the background in units of string length becomes  large, see e.g. \cite{Bonelli:2003kh, Bonelli:2003zu, Gaberdiel:2018rqv, Eberhardt:2018ouy, Sundborg:2000wp}. (2) When the energy of typical string states/configurations are large compared to the string tension. The latter {may} happen when we are close to the Hagedorn phase transition \cite{Hagedorn:1965st, Lizzi:1989iy} or when strings are probing regions of the background with large redshift factors, e.g. strings probing cosmological or black hole horizons \cite{Lousto:1996hg, deVega:1994hu, Bagchi:2022nvj, Bagchi:2023cfp, Bagchi:2024rje}. Here, we are interested in the latter case where effectively string worldsheet becomes a null surface, hence null strings. As discussed, every null string is necessarily tensionless, but not all tensionless strings are null strings. In case (1), string  worldsheet is still a Lorentzian 2D manifold with Weyl + diffeomorphism as gauge symmetry. 

The worldsheet theory of null strings is, however, a 2D Carrollian field theory. The action for this theory, as famously shown by Isberg-Lindstr\"om-Sundborg-Theodoridis (ILST) \cite{Isberg:1992ia, Isberg:1993av}, may be formally obtained through $T\to 0$ limit of ordinary tensile strings. This action explicitly exhibits Weyl + diffeomorphism symmetry and has been the basis for the null string literature, so far; see e.g. \cite{Schild:1976vq, Gamboa:1989zc, Gamboa:1989px, Sundborg:1994py, Bagchi:2019cay, Bagchi:2020fpr, Bagchi:2020ats, Bagchi:2021ban}; see \cite{Bagchi:2026wcu} for a recent review and references therein. Nonetheless, as we pointed out in a recent short note \cite{Sheikh-Jabbari:2026cnj}, the ILST action enjoys invariance under a codimension-1 gauge symmetry, i.e. a gauge symmetry generator is a function of a single direction on the 2D worldsheet. Presence of this gauge symmetry has profound consequences on the null string analysis and puts into question the consistency of the analyses, and thus the results, discussed in the null string literature. 

In this note, we show how the codimension-1 gauge symmetry of \cite{Sheikh-Jabbari:2026cnj} can be promoted to a standard gauge symmetry. To this end, we recall the null string worldsheet is a 2D Carrollian geometry which allows for two distinct Weyl scalings, the \textit{Carroll-Weyl scalings}, this is in contrast to the single Weyl scaling available in the standard Lorentzian geometries. The null string theory is then naturally defined as the action that realizes the two Carroll-Weyl, and the diffeomorphisms as gauge symmetry. This null string action has a Carroll-Weyl gauge field. Nonetheless, as we show after fixing the Carroll-Weyl scaling, the action reduces to ILST action. Moreover, we show that the residual symmetry after this gauge fixing is precisely the codimension-1 gauge symmetry of the ILST action, discussed in \cite{Sheikh-Jabbari:2026cnj}. 

\section{Geometric objects in Carrollian geometry}
\label{sec:carrollian_2d_geometry}

Before constructing the action for the null string, we define the geometric data on a 2D Carrollian worldsheet. For introduction to and general discussion on Carrollian geometries  see \cite{Ciambelli:2019lap, Adami:2024rkr, Ciambelli:2025unn}. All objects introduced in this section are defined intrinsically on the worldsheet manifold $\Sigma$, independently of any specific physical theory. 

To keep things general, consider {$p+1$} dimensional smooth Carrollian manifold. It is defined by a \textit{degenerate metric} $h_{ab}, a,b=1,2,\cdots, p+1$, which is positive semidefinite with rank $p$ everywhere ($h_{ab}$ has $p$ nonzero eigenvalues and one vanishing eigenvalue) and the \textit{kernel vector}, a nowhere-vanishing vector field $v^a$ spanning the kernel of $h_{ab}$, i.e.,
\begin{equation}
v^a h_{ab} = 0 .
\end{equation}
$v^a$ is the eigenvector associated with the vanishing eigenvalue. 
The degenerate metric can be expanded using $p$ frame fields $e^I_a$, 
\begin{equation}
    h_{ab}= \delta_{IJ} e^I_a e^J_b, \qquad v^a e^I_a=0,\qquad I, J=1,2, \cdots, p
\end{equation}

While the minimal requirement to define the Carrollian geometry is $(h_{ab}, v^a)$, it is usually endowed with further structures:
\paragraph{Clock form,}  a $1$-form $n_a$ satisfying the normalization condition
\begin{equation}
v^a n_a = 1 .
\end{equation}
The clock form $n_a$ is not unique: any shift $n_a \to n_a + \alpha_I e^I_a$ preserves the normalization $v^a n_a = 1$. $n_a$ defines an Ehresmann connection of the {ruled or supplemented} Carrollian geometry \cite{Ciambelli:2019lap, Ciambelli:2025unn}: its kernel $\ker(n_a)$ specifies a horizontal subspace complementary to the vertical direction spanned by $v^a$. This splits the tangent bundle as $T\Sigma = \mathrm{span}\{v^a\} \oplus \ker(n_a)$. The Ehresmann connection is intrinsic to the  {ruled or supplemented} Carrollian structure and should not be confused with the Weyl connection ${\cal W}_a$ introduced in the next section, which gauges certain Carroll-Weyl scaling of the embedding fields.

\paragraph{Dual frame fields $e^a_I$.} Using the clock field $n_a$, one can define ``inverse frame fields'' $e^a_I$ as the set of all vector fields satisfying
\begin{equation}
n_a e^a_I = 0 , \qquad {e^a_I e_a^J=\delta_I^J,\qquad \delta^a_b=v^an_n+e^a_I e_a^J \qquad \forall I, J}.
\end{equation}

\paragraph{Volume form.} Using the dual frame fields and the clock form one can define the volume form
\begin{equation}
\epsilon_{a_1a_2\cdots a_{p+1}} := \epsilon_{I_1I_2\cdots I_{p}}\ e^{I_1}_{[a_1} e^{I_2}_{a_2}\cdots e^{I_{p}}_{a_{p}}\ n_{a_{p+1}]}.
\label{eq:volume_form}
\end{equation}
where square-brackets denote anti-symmetrization. 
$\epsilon_{a_1a_2\cdots a_{p+1}}$ is non-degenerate and satisfies 
\begin{equation}
    \epsilon_{a_1a_2\cdots a_{p+1}} v^{a_{p+1}} = \epsilon_{I_1I_2\cdots I_{p}}\ e^{I_1}_{a_1} e^{I_2}_{a_2}\cdots e^{I_{p}}_{a_{p}},\qquad \epsilon_{a_1a_2\cdots a_{p+1}} e^{a_1}_{I_1}=\epsilon_{I_1I_2\cdots I_{p}}\  e^{I_2}_{[a_2}\cdots e^{I_{p}}_{a_{p}} \ n_{a_{p+1}]}
\end{equation}

\paragraph{Summary of geometric data for the 2D case ($p=1$).}  The Carrollian geometry of the worldsheet is completely specified by the following data:
\begin{table}[h]
\centering
\begin{tabular}{|c|c|c|}
\hline
\textbf{Object} & \textbf{Notation} & \textbf{Defining Relation} \\
\hline
Degenerate metric & $h_{ab}$ & $h_{ab} = \ell_a \ell_b$ \\
Kernel vector & $v^a$ & $v^a h_{ab} = 0$ \\
Clock form & $n_a$ & $v^a n_a = 1$ \\
Spatial form & $\ell_a$ & $h_{ab} = \ell_a \ell_b$ \\
Dual spatial vector & $\ell^a$ & $n_a \ell^a = 0,\; \ell_a \ell^a = 1$ \\
Volume form & $\epsilon_{ab}$ & $\epsilon_{ab} = n_a \ell_b - n_b \ell_a$ \\
\hline
\end{tabular}
\caption{Summary of geometric objects on a 2D Carrollian worldsheet.}
\label{tab:carrollian_objects}
\end{table}\\
These objects satisfy the completeness relation
\begin{equation}
\delta^a_b = v^a n_b + \ell^a \ell_b ,
\label{eq:completeness}
\end{equation}
which encodes the decomposition of any tangent vector into its components along the null direction $v^a$ and the spatial direction $\ell^a$.
We close this part by a remark on inverse metric. While $h_{ab}$ is degenerate and does not possess an inverse in the usual sense, one can define a \textit{generalized inverse} $h^{ab}$ satisfying $h^{ac} h_{cb} h^{bd} = h^{ad}$. A natural choice consistent with the Carrollian worldsheet  is $h^{ab} = \ell^a \ell^b$. This generalized inverse is not unique; the ambiguity reflects the freedom in the choice of $\ell^a$ up to the transformation $\ell^a \to \tilde{\ell^a}=\ell^a + \beta v^a$, which leaves {the form of} $h^{ab}$ invariant, {i.e. $h^{ab}\to \tilde{\ell}^a\tilde{\ell}^b$}.

\paragraph{Diffeomorphisms.} In any manifold, with any metric structure, the standard Lorentzian/Euclidean geometry or Carrollian alike, one can span the manifold by a set of coordinates, this can be done patchwise. So, if we deal with a 2D geometry with coordinates $\sigma^a$, one is free to define the coordinates up to diffeomorphisms, 
\begin{equation}\label{2D-diffeos}
    \sigma^a \to \sigma^a+ \xi^a(\sigma^b).
\end{equation}
Under diffeomorphisms the Carrollian triple $(h_{ab}, v^a, n_a)$ transform as a two-tensor, a vector and a one-form, respectively.

\section{The two Carroll-Weyl scalings}

In this section we focus on the 2D case of our interest described by the triple $(\ell_a, v^a, n_a)$. In a standard Lorentzian/Euclidean geometry with metric $\gamma_{ab}$, we have the Weyl scaling:
\begin{equation}\label{Weyl-scaling-standard}
    \gamma_{ab} \to e^{2\phi(\sigma)}\ \gamma_{ab},
\end{equation}
while not changing the coordinate patches. In the Carrollian case, however, given the Carrollian triple structure, we have two options for the scalings.  We now have the option of scaling these independently, i.e.\footnote{It is apparent that the two Carroll-Weyl scalings naturally extend to generic $p+1$ dimensional Carrollian manifolds as $ n_a \to e^{\chi_t(\sigma)} n_a,\quad  v^a\to e^{-\chi_t(\sigma)} v^a, e^I_a \to e^{\chi_s(\sigma)} e^I_a$. }
\begin{align}
n_a \to e^{\chi_t(\sigma)} n_a,&\quad \quad v^a\to e^{-\chi_t(\sigma)} v^a\\
\ell_a \to e^{\chi_s(\sigma)} \ell_a,& \qquad {\ell^a \to e^{-\chi_s(\sigma)} \ell^a}.
\end{align}
These transformations affect the derived geometric objects as follows:
\begin{equation}\label{chis-chit}
\begin{split}
h_{ab} = \ell_a \ell_b \to \ e^{2\chi_s} h_{ab},&\qquad {h^{ab} \to \ e^{-2\chi_s} h^{ab}} \\
\epsilon_{ab} = n_a \ell_b - n_b \ell_a &\to e^{\chi_t + \chi_s} \epsilon_{ab}.    
\end{split}\end{equation}

To compare the two Carrollian Weyl scalings with the standard Weyl scaling, one can consider different decompositions. In the ordinary 2D manifolds, the volume form and the metric both scale as $e^{2\phi}$. Therefore, one may identify the \textit{volume-modulating scaling},
defined by the symmetric combination:
\begin{equation}
\phi := \chi_t = \chi_s.
\end{equation}
Under this transformation,
\begin{equation}\label{phi-scaling}
n_a \to e^{\phi} n_a, \qquad v^a\to e^{-\phi} v^a,\qquad \ell_a \to e^{\phi} \ell_a, \qquad 
\epsilon_{ab} \to e^{2\phi} \epsilon_{ab}, \qquad h_{ab} \to e^{2\phi} h_{ab},
\end{equation}
which scales  $\ell_a, n_a$ in the same way. In the 2D case, in the basis where $h_{ab}$ is diagonal, 
\begin{equation}
    h_{ab}=\left(\begin{matrix} 0 & 0 \\ 0 & h  
    \end{matrix}\right)
\end{equation}
we see that ${\cal V}^a$ 
\begin{equation}\label{calVa}
    {\cal V}^a:= {\sqrt[4]{h}}\ v^a,    
\end{equation}
remains invariant under $\phi$-scaling. ${\cal V}^a$ is hence a vector density, i.e. under diffeomorphisms \eqref{2D-diffeos} it transforms as,
\begin{equation}\label{diff-xi-V}
    \delta_\xi \mathcal{V}^a = \xi^b \partial_b \mathcal{V}^a - \mathcal{V}^b \partial_b \xi^a + \frac{1}{2} (\partial_b \xi^b) \mathcal{V}^a.
\end{equation}
${\cal V}^a$ may be compared with $\sqrt{-\gamma} \gamma^{ab}$ in the standard 2D geometry, which is invariant under the standard Weyl scaling \eqref{Weyl-scaling-standard}, while behaves as a tensor-density of weight +1 under diffeomorphisms,
\begin{equation}\label{diff-xi-gamma-ab}
    \delta_\xi (\sqrt{-\gamma} \gamma^{ab}) = \sqrt{-\gamma}\left(-\nabla_a \xi_b-\nabla_b \xi_a+  (\nabla_c \xi^c) \gamma_{ab}\right),
\end{equation}
where $\xi_a=\gamma_{ab} \xi^b$ and $\nabla_a$ is covariant derivative w.r.t $\gamma_{ab}$.

One may then note that we also have \textit{volume-preserving scaling},
\begin{equation}
\chi := \chi_t = -\chi_s.
\end{equation}
Under this transformation,
\begin{equation}\label{chi-sclaing}
\begin{split}
n_a \to e^{\chi} n_a, \qquad &v^a \to e^{-\chi} v^a, \qquad \ell_a \to e^{-\chi} \ell_a, \\
{\cal V}^a\to e^{-2\chi} {\cal V}^a,
\qquad 
\epsilon_{ab} &\to e^{\chi - \chi} \epsilon_{ab} = \epsilon_{ab}, \qquad h_{ab} \to e^{-2\chi} h_{ab}.   
\end{split}\end{equation}
As the name suggests, the volume form $\epsilon_{ab}$ remains invariant. The volume-preserving scaling is the key new option available in the Carrollian case, which has no counterpart in the standard manifolds. In the next section we study consequences of this new possibility for the null string theory.

\section{Null string action with gauged Carroll-Weyl scalings}

String worldsheet theory (the Polyakov action) may be defined as the action for $D$ target space coordinates $X^\mu$ and the worldsheet metric $\gamma_{ab}$ which has 2D diffeomorphisms+Weyl scaling as the local gauge symmetry \cite{Green:1987sp, Polchinski:1998rq}. In a similar manner, one may \textit{define} null string theory as the action for $X^\mu$ and Carrollian worldsheet objects $n_a, v^a, \ell_a$ which exhibits diffeomorphisms+the two Carroll-Weyl scalings as gauge symmetry. 

To write such an action, let us start with the  standard ILST null string action \cite{Isberg:1993av},
\begin{equation}\label{ILST-action}
S_{\text{ILST}} = \frac{\kappa}{2} \int d^2\sigma \, \mathcal{V}^a \mathcal{V}^b \, \partial_a X^\mu \partial_b X^\nu \eta_{\mu\nu},
\end{equation}
where $X^\mu(\tau,\sigma)$, $\mu = 0,1,\dots, D-1$. In the above, ${\cal V}^a$ is defined in \eqref{calVa} and is invariant under the $\phi$ scaling \eqref{phi-scaling}: explicitly, 
\begin{equation}
    \delta_\phi {\cal V}^a= 0, \qquad \delta_\phi X^\mu=0
\end{equation}
is clearly a gauge symmetry of the above action.\footnote{Recall that in the usual Polyakov action, we have $\sqrt{-\gamma} \gamma^{ab}$ which is invariant under the standard Weyl scaling \eqref{Weyl-scaling-standard}.} 

The ILST action \eqref{ILST-action} is not invariant under the $\chi$ Carroll-Weyl scaling \eqref{chi-sclaing}. To make it invariant we need to replace $\partial_a$ with a covariant derivative $D_a$ and introduce a gauge field, the \textit{Carroll-Weyl connection} ${\cal W}_a$ associated with the $\chi$-scaling. Explicitly,\footnote{See \cite{Gustafsson:1994kr} where a similar action, though in a slightly different context and different physical analysis and results,  had appeared.}
\begin{equation}\label{gauged-ILST-action}
S = \frac{\kappa}{2} \int d^2\sigma \, \mathcal{V}^a \mathcal{V}^b \, D_a X^\mu D_b X^\nu \eta_{\mu\nu}, %\qquad D_a = \partial_a + {\cal W}_a,
\end{equation}
restores the full Carroll-Weyl invariance, where
\begin{equation}
D_a X^\mu := \partial_a X^\mu+ {\cal W}_a X^\mu,
\end{equation}
with,
\begin{equation}\label{chi-transf}
X^\mu \to e^{2\chi} X^\mu, \qquad \mathcal{V}^a \to e^{-2\chi} \mathcal{V}^a, \qquad {\cal W}_a \to {\cal W}_a-2\partial_a \chi, 
\end{equation}
and $\chi(\tau,\sigma)$ is an arbitrary scalar function. As the above shows, and recalling \eqref{chi-sclaing}, the Carroll-Weyl connection ${\cal W}_a$ is \textit{not} the Ehresmann connection $n_a$. 
Equation \eqref{gauged-ILST-action}, similarly to the ILST action \eqref{ILST-action}, is clearly also invariant under worldsheet diffeomorphisms:
\begin{equation}\label{generic-diff}
\begin{split}
\delta_\xi X^\mu &= \xi^a \partial_a X^\mu, \\
\delta_\xi \mathcal{V}^a &= \xi^b \partial_b \mathcal{V}^a - \mathcal{V}^b \partial_b \xi^a + \frac{1}{2} (\partial_b \xi^b) \mathcal{V}^a, \\
\delta_\xi {\cal W}_a &= \xi^b \partial_b {\cal W}_a + {\cal W}_b \partial_a \xi^b . 
\end{split}\end{equation}

In summary, the action \eqref{gauged-ILST-action} is the appropriate null string action with all expected gauge symmetries, 2D diffeomorphisms and the two $\phi, \chi$ Carroll-Weyl scalings. Explicitly, and in the infinitesimal form, for a generic symmetry transformation parameter $\eta:=(\xi^a, \chi, \phi)$,
\begin{subequations}\label{generic-transf-eta}
\begin{align}
\delta_\eta X^\mu &= \xi^a \partial_a X^\mu + 2\chi X^\mu, \label{eq:eta-X}\\
\delta_\eta \mathcal{V}^a &= \xi^b \partial_b \mathcal{V}^a - \mathcal{V}^b \partial_b \xi^a + \frac{1}{2} (\partial_b \xi^b) \mathcal{V}^a-2\chi {\cal V}^a, \label{eq:eta-V}\\
\delta_\eta {\cal W}_a &= \xi^b \partial_b {\cal W}_a + {\cal W}_b \partial_a \xi^b -2\partial_a\chi . \label{eq:eta-A}
\end{align}\end{subequations}
the action remains invariant. 

The dynamical fields of the action \eqref{gauged-ILST-action} are $X^\mu, {\cal V}^a$ and the gauge field ${\cal W}_a$. That is, we have $D+2+2$ fields which are functions of the two variables $\sigma^a$. There are $2+1$ gauge transformations; we have not counted the inert $\phi$-scaling, as none of the fields have a nonvanishing variation under $\phi$. We end this section with {some comments regarding the Carroll-Weyl connection ${\cal W}_a$ and the gauged action \eqref{gauged-ILST-action}: (1) As it is explicitly seen, the action only involves component of the gauge field along the kernel vector, i.e. ${\cal V}^a {\cal W}_a$ combination. The other component, $\ell^a {\cal W}_a$ does not appear in the action and hence in the classical description of the theory. Therefore, we indeed have $D+2+1$ degrees of freedom, plus $2+1$ gauge symmetries.} (2) Due to Carrollian nature of the theory, ${\cal V}^a, {\cal W}_a$ do not have standard kinetic terms, and both of them can be gauge-fixed, leaving to propagating d.o.f. (3) In the Carrollian case we do not have the option of a $\chi$-invariant  kinetic term for the gauge field ${\cal W}_a$; the canonical choice ${\cal V}^a{\cal V}^b h^{cd} F_{ac}F_{bd}$ with $F_{ab}:=\partial_{[a}{\cal W}_{b]}$, does not work because of the $\chi$-scalings of ${\cal V}^a$ and $h^{ab}$, cf. \eqref{chi-sclaing}.  In a similar way, one can show that one cannot construct a scalar under diffeomorphism from any combination of ${\cal V}^a, h^{ab}, F_{ab}$ that is $\chi$-scaling invariant.\footnote{{Note that with the same dynamical contents, we still have the option of adding a topological term $\epsilon^{ab} F_{ab}$. However, this term does not have any $X^\mu$ dependence and being a total derivative no bearing on the classical theory.}} So, \eqref{gauged-ILST-action} is {a fairly general string theory action} that is $\chi$-scaling and diffeomorphism invariant.

\section{Relation between gauged null string and ILST actions}

As pointed out, the ILST action \eqref{ILST-action}, while diffeomorphism invariant, is not invariant under $\chi$-scaling. The gauged action, on the other hand, has a larger gauge symmetry plus one more dynamical degree of freedom ${\cal W}_a$. Recalling the Carrollian form of the action, one may relate the two actions by a gauge fixing. The gauge-fixing condition  ought to be diffeomorphism invariant. Therefore, one may {locally} fix the gauge
\begin{equation}\label{chi-gauge-fixing}
    {\cal V}^a {\cal W}_a=0 
\end{equation}
To explore the residual gauge symmetries after the above gauge-fixing, we note that
\begin{equation}
    \delta_\chi ({\cal V}^a {\cal W}_a)= -2{\cal V}^a (\partial_a\chi+ {\cal W}_a \chi)= -2 {\cal V}^a \partial_a\chi
\end{equation}
where in the second equality we again used \eqref{chi-gauge-fixing}. Thus, \eqref{chi-gauge-fixing} fixes the $\chi$-scaling up to $\chi$'s satisfying ${\cal V}^a \partial_a\chi=0$. The residual symmetry after \eqref{chi-gauge-fixing} is therefore, functions $\chi$ that are defined on codimension-1 surfaces over the worldsheet.

Let us now rewrite the gauged action \eqref{gauged-ILST-action} in the gauge \eqref{chi-gauge-fixing}. In this gauge ${\cal V}^a D_a X^\mu={\cal V}^a\partial_a X^\mu$ and  we recover the ILST action \eqref{ILST-action}. The above discussion sheds light on the analysis in \cite{Sheikh-Jabbari:2026cnj}: The ILST action still exhibits invariance under the residual set of $\chi$-scalings that satisfy ${\cal V}^a \partial_a\chi=0$. 

%%%%%%%%%%%%%%%%%%%%%%%%%%%%%%%%%%%%%%%%%%%%%%%%%%%%%%%%%%%%%%%%%%%%%%%%%%%%%%%%%%%%
\section{Equations of motion of gauged null string action}

The action \eqref{gauged-ILST-action} has some dynamical and some gauge degrees of freedom. For a full analysis of the theory we need to analyze equations of motion (EoM) and the gauge-fixing conditions and the residual gauge symmetries. While a full analysis is left to a separate publication, here we briefly discuss the EoM that are complementary to gauge-fixing conditions. Varying the action with respect to $X^\mu$, $\mathcal{V}^a$, and ${\cal W}_a$ independently yields the following field equations.

\paragraph{EoM for $X^\mu$:}
After integrating by parts and discarding boundary terms,
\begin{equation}
D_a\!\left( \mathcal{V}^a \mathcal{V}^b D_b X_\mu \right)= \partial_a \left( \mathcal{V}^a \mathcal{V}^b D_b X_\mu \right) - {\cal W}_a \mathcal{V}^a \mathcal{V}^b D_b X_\mu = 0.
\label{eq:eom_X}
\end{equation}
This is the covariant conservation of the current $J^a_\mu := \mathcal{V}^a \mathcal{V}^b D_b X_\mu$, reflecting that $X^\mu$ carries charge $+2$ under the scaling symmetry. If 
\begin{equation}
    P_\mu:= \kappa \mathcal{V}^a D_a X_\mu,
\end{equation}
the above equation may be written as 
\begin{equation}
    D_a ({\cal V}^a P^\mu)=0 
\end{equation}

\paragraph{EoM of $\mathcal{V}^a$:}
\begin{equation}
{\cal C}_a:= \frac{\delta S}{\delta \mathcal{V}^a} = \kappa \; \mathcal{V}^b \, D_a X^\mu D_b X_\mu = D_a X^\mu P_\mu =0 \qquad (a = 0,1),
\label{eq:eom_V}
\end{equation}
i.e., $P\cdot D_a X = 0$ for each $a$. Since ${\cal V}^a$ have vanishing momentum, they are gauge degrees of freedom and hence ${\cal C}_a=0$ are constraints. 

\paragraph{EoM of ${\cal W}_a$:}
Using $\partial (D_b X^\mu)/\partial {\cal W}_a = \delta_b^a X^\mu$, we obtain
\begin{equation}
\frac{\delta S}{\delta {\cal W}_a} = \kappa \; \mathcal{V}^a \mathcal{V}^b X^\mu D_b X_\mu = {\cal V}^a\ {\cal C}_3=0, \qquad {\cal C}_3:= P\cdot X.
\label{eq:eom_A}
\end{equation}
${\cal C}_3$ is also a gauge constraint, since ${\cal W}_a$ has no kinetic term (and has vanishing momentum). 

\paragraph{EoM in ${\cal V}^a {\cal W}_a=0$ gauge.} In this gauge, $P^\mu=\kappa\ {\cal V}^a \partial_a X^\mu$ and $D_a {\cal V}^a=\partial_a{\cal V}^a$. Thus, EoM reduce to, 
\begin{subequations}
\begin{align}
       \partial_a ({\cal V}^a P^\mu) &=\partial_a({\cal V}^a {\cal V}^b \partial_b X^\mu)=0 \\ 
       {\cal C}_v:= {\frac{1}{\kappa}} {\cal V}^a {\cal C}_a&= P\cdot P =0 \\ 
        {\cal C}_l:= {\ell}^a {\cal C}_a&= P\cdot ({\ell}^a \partial_a X)+{\ell}^a {\cal W}_a\ {\cal C}_3=P\cdot ({\ell}^a \partial_a X)=0 \\
        {\cal C}_3&= P\cdot X=0
\end{align}
\end{subequations}
We crucially note that in ${\cal C}_l$ the remaining component of ${\cal W}_a$, ${\ell}^a {\cal W}_a$ (that is not fixed by the gauge-fixing \eqref{chi-gauge-fixing}) is proportional to ${\cal C}_3$ and hence drops out. So, ${\cal W}_a$ has no classical dynamics and dynamically and at classical level the action \eqref{gauged-ILST-action} is equivalent to ILST action \eqref{ILST-action}. A more detailed analysis of finding physical null string configurations, i.e. solving EoM subject to constraints and fixing the residual symmetries, will be presented in a companion paper \cite{Sheikh-Jabbari:2026tpf}. 

\section{Discussion and outlook}

The purpose of this work was to put the overlooked codimension-1 $\chi$ symmetry of the ILST action \cite{Sheikh-Jabbari:2026cnj} on a firm geometric and physical ground. We achieved this by a careful analysis of the options Carrollian geometries provide. Specifically, we noted that Carroll geometry provides two Carroll-Weyl scalings and the proper null string action should exhibit both as gauge symmetries. The ILST action is a gauge-fixed form of our null string action with Carroll-Weyl gauge symmetry.

It is customary to obtain Carrollian theories as $c\to 0$ limit of Lorentzian theories. In the null string case, this also amounts to taking the tensionless limit \cite{Sundborg:1994py, Bagchi:2019cay, Bagchi:2020fpr, Bagchi:2020ats, Bagchi:2021ban, Bagchi:2026wcu}. This yields the ILST action, with the  Carroll-Weyl scaling symmetry generated by $\chi$, limited to ${\cal V}^a\partial_a\chi=0$, the gauge-fixed form of our action \eqref{gauged-ILST-action} \cite{Sheikh-Jabbari:2026cnj}. One can ask if it is possible to obtain the gauged action \eqref{gauged-ILST-action} through the tensionless/Carrollian limit of the {textbook tensile string theory}, perhaps in a more careful limit procedure. Our discussions in this paper clarify that the answer is No: The option of two Carroll-Weyl scalings and in particular the volume-preserving Carroll-Weyl scaling is only available for Carrollian geometries, with no counterpart in the Lorentzian worldsheet. This symmetry enhancement appears only at the Carrollian point and not in $c\ll 1$ worldsheet. 

In this short note, we {have} focused on introducing the Carroll-Weyl scaling gauged action. This action is the appropriate starting point of reloading the null string literature.  A thorough analysis of the classical theory and its subsequent quantization of the null string theory are the obvious next steps to be explored in upcoming publications. In particular, in two companion papers we complete the classical analysis by working through the Hamiltonian analysis of gauged systems for the null strings \cite{Sheikh-Jabbari:2026tpf} and first steps of quantization are carried out in \cite{Rasulian:2026jvg}. We close by the comment that the Carroll-Weyl scaling also exists for null $p$-branes and the analysis for null branes, including the ones in \cite{Dutta:2024gkc}, should be thoroughly revisited too.

\begin{acknowledgments}
We thank Arjun Bagchi, Aritra Banerjee, Daniel Grumiller and Ida Rasulian for discussions and Giulio Bonelli, Ulf Lindstorm and Bo Sundborg for helpful email exchanges and discussions. MMShJ acknowledges Iranian National Science Foundation (INSF) research chair grant No.40451653. The work of HY is supported in part by Beijing Natural Science Foundation under Grant No. IS23013.
\end{acknowledgments}

\end{document}